\documentclass[%
twocolumn,%
showpacs,%
showkeys,%
preprintnumbers,%
amsmath,amssymb,%
page%
]{revtex4}
\usepackage{graphicx}% Include figure files
\usepackage{dcolumn}% Align table columns on decimal point
\usepackage{bm}% bold math
\def\timeevol#1{\setbox1=\hbox{$\longmapsto$}\setbox2=\hbox{$
    \scriptstyle #1$}\copy1\kern-.5\wd1\kern-.5\wd2
    \raise-1.2\ht2\copy2\kern-.5\wd2\kern\wd1}
\begin{document}

%\preprint{Final Draft twseg4 7/19/05}
%\preprint{Draft (twseg3), 7/14/2005}

\title{
Metastable Congested States in Multisegment
Traffic Cellular Automaton}
\author{
Yutaka Nishimura${ }^{1}$, Taksu Cheon${ }^{1}$  and Petr \v Seba${
}^{2}$ }

\affiliation{
${ }^1$
Laboratory of Physics, Kochi University of Technology,
Tosa Yamada, Kochi 782-8502, Japan\\
${ }^2$
Department of Physics, University of Hradec Kralove,
%Vita Nejedleho 573,
Hradec Kralove CZ50003, Czech Republic
}

\date{November 8, 2005}

\begin{abstract}
We investigate a  simple multisegment  cellular automaton model of traffic flow.
With the introduction of segment-dependent acceleration probability, 
metastable congested states in the intermediate density region emerge, 
and the initial-state dependence of the flow is observed.
The essential feature of three-phased structure empirically found
in real-world traffic flow is reproduced without elaborate assumptions.
\end{abstract}


\maketitle
%

% % % % % % % % % %
% % % % % % % % % %
\section{Introduction}
The cellular automaton model has established itself as
a standard in the analysis of traffic flow
with its simplicity, affinity to computer simulations, and
also with its extendibility to handle the realistic traffic \cite{NS92, SS95}.
A particularly appealing aspect is
its capability to capture the essential characteristics of the traffic flow
with very few system parameters, which is a surprising economy of assumptions.
% sometimes even with analytical results.
%
With further assumptions for the adoption to the specific cases,
the model is shown to be capable of reproducing down to 
the fine details of traffic flow
in various settings, and it has been used for the analysis of real-life traffic
with great success in past decade \cite{NA97}.

Lost in this process of elaboration, however, is the
elegance of generality which the original standard model possesses.
There is an obvious generic feature of the traffic flow
that is beyond the reach of original model, but appears to await the
explanation with a simple model.
That is the three phase structure.
In a typical traffic, the low and high density regions show the free motion and
traffic jam.
In the fundamental diagram of traffic, that plots the flux against the density,
they are respectively represented by the increasing and decreasing lines,
both of which are reproduced well with the standard model.
In the intermediate density, however, empirical data indicate the
existence of the third phase, in which there is no clear single line
in the fundamental diagram, but the points are scattered in broad region,
indicating the existence of the fluctuation in the traffic flow.

In this article, we intend to show that this third phase appears in
a very simple extension of the standard model that includes segments
which has different system parameters.
These segments are introduced to represent the natural bottlenecks
-- narrow road segments, hilly segments, a segment under construction, {\it etc.} --
that are ubiquitous in real-life traffic.
We base our discussion primarily on numerical simulations and supplement them
with simple intuitive physical arguments.

It has been recognized for some time that the fluctuation in the traffic flow 
is closely related to the
the appearance of the metastable states \cite{BS98, AS01, NI04}.
We find that the emergence of the third phase is also 
a direct result of the metastable states that appear 
with a minimal assumption in our model.

% % % % % % % % % %
% % % % % % % % % %
\section{Traffic Model
with Probabilistic Acceleration in Multi-Segmented Road}

We first lay out our model.
A road is divided into a one-dimensional array of $L$ cells  
on which $M$ cars are placed.
The location ({\it i.e.} cell number) and the velocity of the cars
are represented by integers $x_i$ and $v_i$ with the car index $i=1,2, ..., M$. Cars
are indexed in ascending order to the direction of the motion
without overlap; $x_1 < x_2 < ... < x_M$.
The road is considered to be circular, so that the $(L+1)$-th cell is identified with the fist cell.
The road is subdivided into
$S$ segments ${\cal S}_1=$ $\{ x | 0 \!<\! x \le L_1 \}$,
${\cal S}_2=$ $\{ x | L_1\! <\! x \le L_1\!+\!L_2 \}$, ...,
${\cal S}_S=$ $\{ x | L_1+...+L_{S-1}\! <\! x \le L \}$,
with the constraint on the sub-length $L_1+...+L_S = L$.
%Choice of $L_a$ is arbitrary since the cyclic boundary is used.
Each segment, $s=1,...,S$, is assigned its own maximum velocity $U_s$ and
non-acceleration probability $R_s$.  
The time evolution of the system in discrete time steps $t \to t+1$
is described by the following three consecutive updating for $v_i$
and $x_i$ $\in S_s$;
% 1
\begin{eqnarray}
\label{rule}
& &{\rm 1)\  Probabilistic\  acceleration:}
\nonumber \\
& & \quad
v_i \longrightarrow {\rm min}(v_i+1 , U_s) {\rm \ with\ probability}  (1-R_s)
\nonumber \\
& &{\rm 2)\ Breaking\ to\ avoid\ collision:}
\nonumber \\
& & \quad
v_i \longrightarrow {\rm min} (v_i , x_{i+1}-x_i-1).
\\
& &{\rm 3)\ Advancement:}
\nonumber \\
& & \quad
x_i \longrightarrow x_i+v_i.
\nonumber
\end{eqnarray}
%
%We assume that
The updating is done in parallel for all cars $i = 1,2, ... M$:
Namely, the current position of the preceding car is used to calculate
the distance $x_{i+1}-x_i-1$.

The model can be thought of as a variant of the standard 
Nagel-Schreckenberg (NS) model
in that it follows the acceleration and deceleration step-by-step.
In fact, if we change the ordering of ``breaking'' and ``random
deceleration'' in the standard NS model, we obtain a model where
cars accelerate, and then randomly decelerate. The model then
becomes almost identical to ours, apart from the treatment of the
cars running at the maximum velocities. This inverting of the
order effectively prohibits the cars running in the jam from
random deceleration, thus models the ``heightened alertness'' of
drivers who have the preceding cars in their close sight.
Our model is also related to the Fukui-Ishibashi (FI) model in that both employ
probabilistic acceleration, which is in contrast to the NS model that treats the
process by splitting it into acceleration and probabilistic deceleration.

Using the second step of the rule (\ref{rule}), we may
define a {\it jam} as a block of cars
in which the distance of neighboring cars is below the maximum velocity,
%Note that there is a mechanism for aggregation of a block of jam.
%That is, a car hitting a jam composed of a block of equidistant cars is added to the jam.
%There also is a competing mechanism of percolation of jam, in which the car at the head of
%the block of jam gets away from the block with the probability ($1-R_i$).
%
A crucial feature of our model is the existence of the 
{\it metastable block of jam} made up of equi-spaced cars.
This is a direct result of our model in which no further random deceleration
occurs for cars inside the jam.
The only limiting factor of the growth of this block is
the percolation at the head of the block whose rate is proportional to $(1-R_s)$.
This feature captures oft observed self-forming bottleneck of slow-moving
cars in single-lane roads in real-life traffic.   This could be regarded as 
a main justification for devising yet another variant of traffic cellular automaton model.
%

%

% % % % % % % % % %
% % % % % % % % % %
\section{Numerical Results}

We perform numerical simulations on the model presented above
with one segment, two-segments, and six segments cases,
the last one being the model of generic realistic traffic.

The macroscopic characteristics of traffic systems are best summarized with
so-called fundamental diagram, which is the traffic flux $F$ plotted as the function of
traffic density $\rho$, each of which are defined as
% 2, 3
\begin{eqnarray}
\rho \!\! &=& \!\! \frac{M}{L} ,
\\
F 
\!\! &=& \!\! 
\rho \frac{1}{M} \left<\right. \sum_{i=1}^{M} v_i \left.\right>_t 
= 
\frac{1}{L}\left<\right. \sum_{i=1}^{M} v_i \left.\right>_t  .
\end{eqnarray}
Here, $\left< \right>_t$ stands for the average over many time
steps $t$.
%

%%%%%%
% FIG 1
%
\begin{figure}
% % %\center\includegraphics[width=14.0cm]{f1.eps}
\center\includegraphics[width=7.0cm]{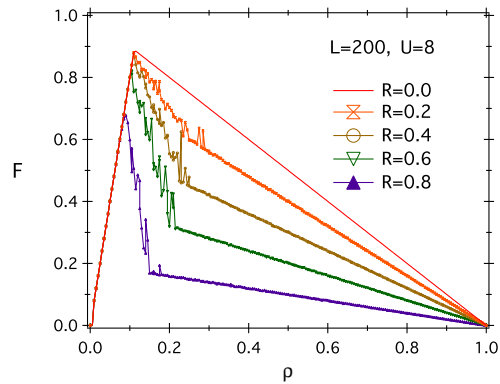}
\caption
{\label{fig1}
The fundamental diagrams of our model.   The non-acceleration rate $R$
is varied from $R=0$ to $0.8$ as indicated in the graph.  The maximum velocity
is fixed at $U=8$ for all cases.  The size of the road is set to be $L = 200$.
}
\end{figure}
We first show the results of the single-segment traffic in FIG. 1.
The size of the road is set to be $L=200$ and the maximum velocity is
chosen to be  $U=8$. Fundamental diagram with various non-acceleration rates
$R$ are shown.
Each point in the graph represents the result
of 20,000 iterations starting from a single random initial configuration.
In the diagram, we observe three distinct regimes of traffic flow.  In between
the free traffic regime at low density and  jamming regime at high density,
there is a distinct 
%partial jamming 
third regime at intermediate density.
Here, the flux displays the sensitive dependence on the initial condition, which is
shown in the graph as jittery structures.  

It is certainly possible to consider
the ensemble of initial conditions among which the results could be averaged
out to obtain smooth curves.  However, we can also argue that jittery structure
reflects a special feature of the model in which
there can be metastable states that result in different flux for traffics with
same density but different initial configurations.
We stress that, at this stage,
the existence of the intermediate region is a result of finite size effect.
This can be easily proven by the explicit numerical calculation 
showing the shrinking of this region with larger $L$.
At the limit $L \to \infty$,  fundamental diagram with two phases
with discontinuous $F(\rho)$ is obtained.
However, this intermediate region develop into a distinct {\it phase}
when there are more than two segments, $S \ge 2$ in a road.
%

%%%%%%
% FIG 2
%
\begin{figure}
% % %\center\includegraphics[width=14.0cm]{f2.eps}
\center\includegraphics[width=7.0cm]{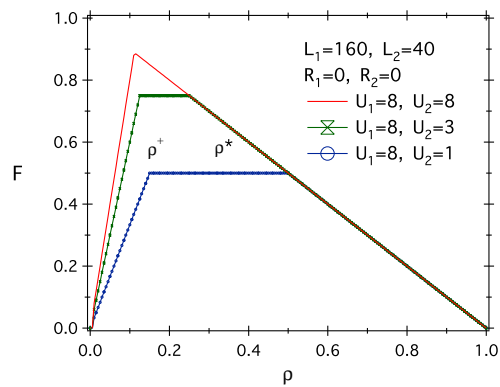}
\caption
{\label{fig2}
The fundamental diagrams of two segment model with a common non-acceleration rates
$R_1 = R_2 = 0$ are shown.  The maximum velocity for the first segment
is fixed at $U_1=8$, and it is varied as $U_2 = 8$, $3$ and $1$ for the second segment.
}
\end{figure}
That fact is clearly shown in FIG. 2, in which the fundamental diagram 
of two segment model
with $L=200$ which is split into $L_1=160$ and $L_2=40$ is shown.
Here, we set a common non-acceleration rate $R_1=R_2=0$ 
and different maximum velocities $U_1$ and $U_2$ for two segments.
We assume the smaller maximum velocity for the second segment,
$U_1 > U_2$. 
%The low velocity section works as a bottleneck that causes congestion to develop
%in the interface region.
%
By the inspection, we sense the existence of analytical solutions represented by
the straight lines connecting $\rho=0$, $\rho=1$ both at $F=0$
and two critical densities $\rho^\dagger$ and $\rho^\star$ both at
the plateau flux $F^\star$.
At the critical  density $\rho^\dagger$, the block of jam with 
$v_i=U_2$ and car spacing $x_{i+1}-x_i=U_2+1$ is formed 
in the second segment while the traffic is at full speed $v_i=U_1$ 
with  car spacing $U_1+1$ in the first segment.
At $\rho^\star$, however, the jam with $v_i=U_2$ fills the entire system. 
With elementary calculations, we obtain the critical densities in the forms
% 4
\begin{eqnarray}
\rho^\dagger &=& \frac{L_1}{L}\frac{1}{U_1+1}+\frac{L_2}{L}\frac{1}{U_2+1} ,
%\end{eqnarray}
%
%and
% 5
\\
%\begin{eqnarray}
\rho^\star &=& \frac{1}{U_2+1} .
\end{eqnarray}
We also obtain the plateau flux
% 6
\begin{eqnarray}
\label{fstr}
F^\star = \frac{U_2}{U_2+1} .
\end{eqnarray}
At the plateau region $\rho^\dagger \le \rho \le \rho^\star$,  the flow is limited by the block
of jam with car spacing $U_2+1$ which now extends into the high maximum velocity region,
in which drivers perceive the block as {\it bottleneck congestion}.
The length of the block is determined by the requirement that ``free traffic'' segment has
the density that corresponds to the same flux with the block, (\ref{fstr}).
Note that the difference between two critical densities $\rho^\dagger$ 
and $\rho^\star$ persists at the continuum  limit $L_s \to \infty$ if the ratio $L_1/L_2$
is kept constant.  This clearly shows that the appearance of the intermediate plateau
region is not the artifact of the discretization nor the finiteness of the model.  Thus this 
third region can be legitimately regarded as the third {\it phase}.
The three phase structure thus obtained is reminiscent to that obtained with
the stop-and-go dynamics \cite{IF01}.
The metastable characteristics of the blocks is essential in obtaining this structure.
%

%%%%%%
% FIG 3
%
\begin{figure}
% % %\center\includegraphics[width=14.0cm]{f3.eps}
\center\includegraphics[width=7.0cm]{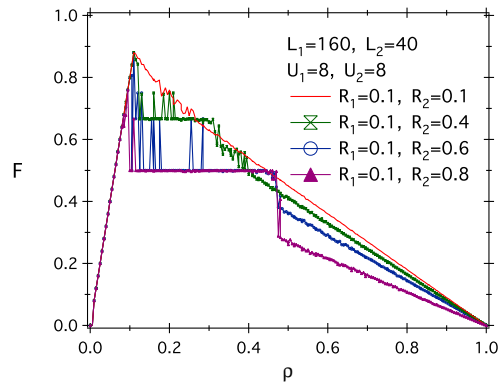}
\caption
{\label{fig3}
The fundamental diagrams of two segment model with a common maximum velocity
$U_1 = U_2 = 8$ are shown. The non-acceleration rates for the first segment
is fixed at $R_1  = 0.1$ and that of second segment is varied as $R_2 = 0.4$,
$0.6$ and $0.8$.
Line connecting the data points are drawn solely to indicate the fluctuation between
discretized  flux: There are no points in-between the value specified by (7) in the
intermediate density region.
}
\end{figure}
%

%%%%%%
% FIG 4
%
\begin{figure}
\center
%{   \includegraphics[width=8.0cm]{f4a.eps}\ 
%    \includegraphics[width=8.0cm]{f4b.eps}
%}
{   \includegraphics[width=4.0cm]{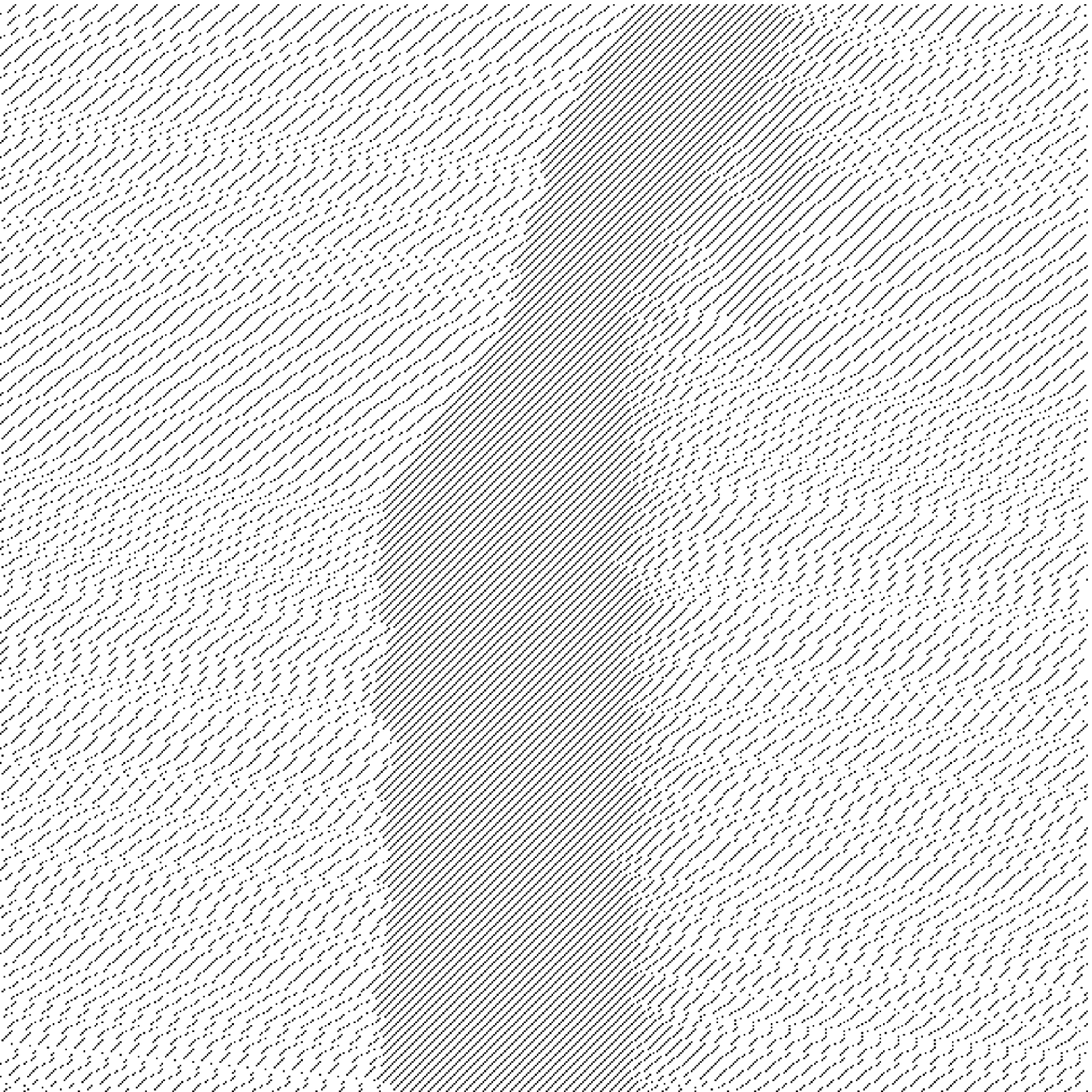}\ 
    \includegraphics[width=4.0cm]{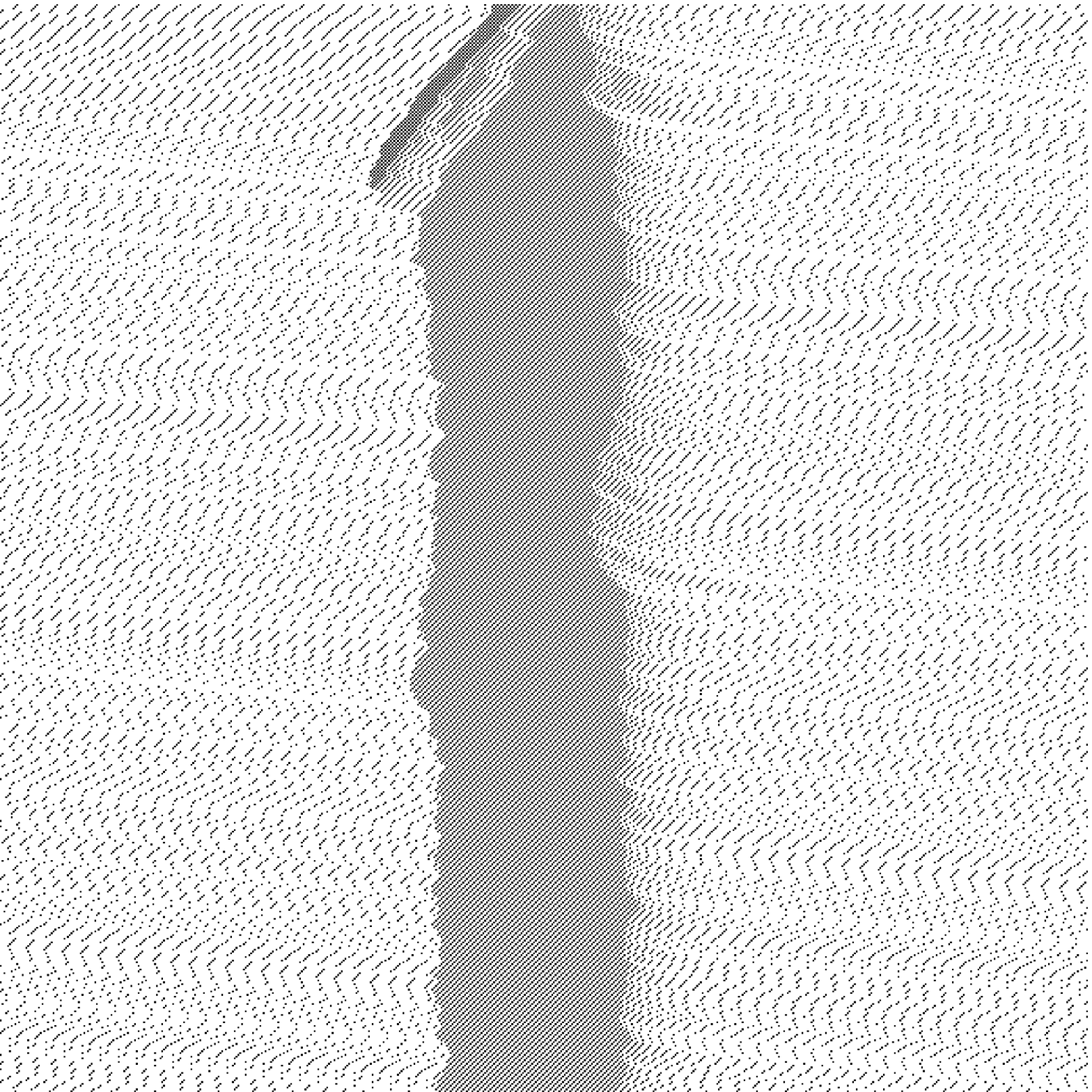}
}
\caption
{\label{fig4}
The flow patterns of two segment model at $\rho=0.3$.  The system parameters are
$U_1 = U_2 = 8$,  $R_1=0.1$ and $R_2=0.5$. 
Different initial conditions result in different metastable states: 
The graph on the left hand side shows the block of jam with $v_i=4$,
while the right hand side shows the one with $v_i=2$.
}
\end{figure}
The situation becomes even more interesting when different non-acceleration rates
are assigned to the different segments.
In FIG. 3, the fundamental diagrams with a common $U_s$ and the
different $R_s$ are shown.
Actual values are chosen to be $U_1=U_2=8$,  $R_1=0.1$ and $U_2=0.6$.
Here, in between the free traffic phase
at low density region $\rho < \rho^\dagger$
and jamming phase at high density region $\rho > \rho^\star$,
we clearly observe an intermediate phase where there is wild oscillations
in $F(\rho)$.  Different outcome is obtained from different initial configuration.
A most striking feature, however, is the fact that, in this intermediate phase, 
the values of the flux $F$ is {\it discretized} at the fractions
% 7 
\begin{eqnarray}
F = \frac{V}{V+1} ,
\quad V=1, 2, ..., U_1 .
\end{eqnarray}
The cause of this discretization and fluctuation is again the
appearance of metastable blocks composed of equi-spaced cars
moving in common velocity $V$. The road segment with high-$R_s$
hinders the percolation of the blocks, and helps the development
of its aggregation. Thus, in effect, the high $R_s$ segment works
very similar to the low-$U_s$ segment in previous example. A
crucial difference, however, is that the block velocity $V$ is not
preset. Metastable blocks with different $V$ can appear and
dominate depending on the initial condition, which results in
non-unique value for $F$ at each $\rho$.

Our claim of the appearance of plural metastable bottleneck blocks are
corroborated by the flow pattern shown in FIG. 4,
which is obtained from the same simulation as in FIG. 3.
FIG. 4(a) and FIG. 4(b) are the results of same parameters but different initial conditions.
clearly, metastable congestion blocks with different internal car spacing (thus different
block velocities) are becoming the limiting factors of the total flow.

%%%%%%
% FIG 5
%
\begin{figure}
% % %\center\includegraphics[width=14.0cm]{f5.eps}
\center\includegraphics[width=7.0cm]{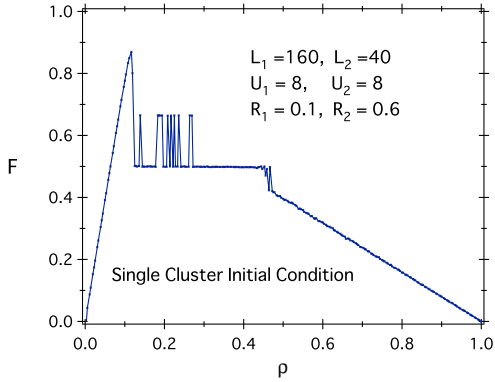}
\caption
{\label{fig5}
The fundamental diagram of two segment model with a common maximum velocity
$U_1 = U_2 = 8$ and different non-acceleration rates $R_1 = 0.1$, $R_2 = 0.6$.
The initial condition is taken to be a single cluster.
}
\end{figure}
%
%%%%%%
% FIG 6
%
\begin{figure}
% % %\center\includegraphics[width=14.0cm]{f6.eps}
\center\includegraphics[width=7.0cm]{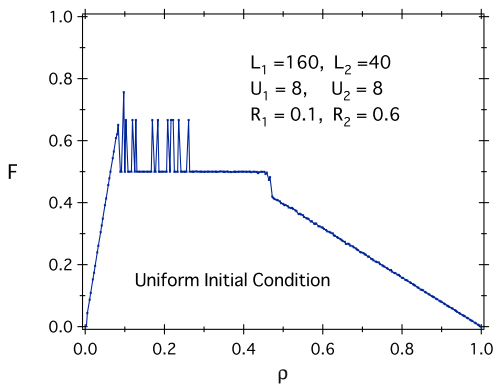}
\caption
{\label{fig6}
Same as FIG. 5 except that the initial condition is taken to be uniformly distributed,
equally spaced cars.
}
\end{figure}
%

%
%
%%%%%%
% FIG 7
%
\begin{figure}
% % %\center\includegraphics[width=14.0cm]{f7.eps}
\center\includegraphics[width=7.0cm]{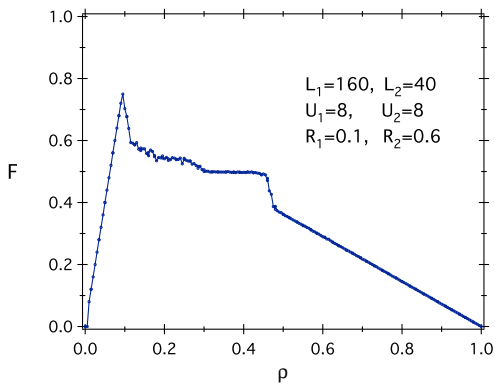}
\caption
{\label{fig7}
The fundamental diagram of two segment model with a common maximum velocity
$U_1 = U_2 = 8$ and different non-acceleration rates $R_1 = 0.1$, $R_2 = 0.6$.
The flux at each density is obtained by averaging over 200 different initial states.
}
\end{figure}

In order to further clarify the initial-condition dependence, we show the results
of two different types of definite initial states:  
In FIG. 5, the fundamental diagram 
of traffic flow calculated from single cluster initial states is shown. 
In FIG. 6, the initial condition is now a uniform distribution, meaning 
equally spaced cars spread over the whole road.  
In both cases, a common maximum velocities $U_1 = U_2 = 8$ and
different non-acceleration rates $R_1 = 0.1$ and $R_2 = 0.6$ are used.
Even with these well-defined non-random initial conditions,
they both show the fluctuating 
flux as the functions of density.  
These diagrams unambiguously show that the  discretization is the results of
random appearance of different metastable states.
In FIG. 7,  the result of average over different initial condition is shown.
The system parameters are the same as those in FIG. 5 and FIG. 6.
The flux now comes in between the discretized value found in 
the case of single initial condition.
This is what we should expect when the fundamental diagram is
drawn from the accumulation of actual data from real-life traffic.

%
%%%%%%
% FIG 8
%
\begin{figure}
% % %\center\includegraphics[width=14.0cm]{f8.eps}
\center\includegraphics[width=7.0cm]{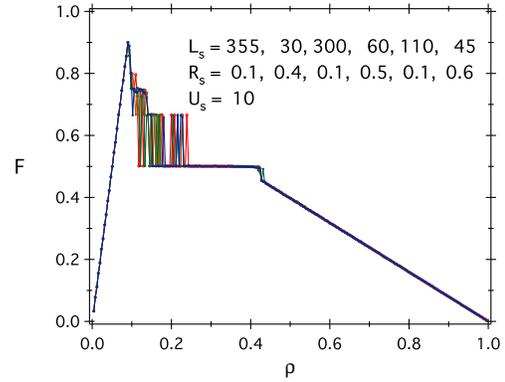}
\caption
{\label{fig8}
The fundamental diagrams of ``generic''  six segment model with a common maximum velocity
$U_1 = ... = U_6 = 8$ are shown. The non-acceleration rates for the odd segments
is fixed at $R_1= R_3 = R_5 = 0.1$, and
that of even  segment is varied as $R_2 = 0.4$, $R_4 = 0.5$, and $R_6 = 0.6$.
The lengths of the segments are $L_1=355$, $L_2=30$, $L_3=300$, $L_4=60$, 
$L_5=110$ and $L_6=45$.
Here, results from a single random initial configuration is shown for 
each value of the density, as in FIG. 3.
}
\end{figure}

We now move on to show that the existence of a phase with fluctuating
discretized flux is not a specific result of the two-segment model,
but a feature generic to  multi-segment traffic in our model.
In FIG. 8, we present the results of six segment traffic, which is brought up as a
toy model of realistic single-lane ``county-road'' traffic.  Road length is set to be
$L=900$.  The road is now made up
of the ``normal'' segments $s=1, 3, 5$ with $R_1=R_3=R_5=0.1$,
and  the bottleneck segments $s=2, 4, 6$, 
each having $R_2=0.4$, $R_4=0.5$ and $R_6=0.6$.
The length of each segments are set to be
$L_1= 355$, $L_2=30$, $L_3=300$, $L_4=60$,  $L_5=110$ and $L_6=45$.
These arbitrary numbers are chosen to make this case a generic example.
A common maximum velocity $U_s=10$ is used for all $s$.
We find the result to be essentially indistinguishable to the case of two segments.
In a sense, this is to be expected because
if several jamming blocks are formed on a road with many segments,
the one with smallest flux among them will eventually absorb the rest
and becomes the limiting bottleneck,
thus determine the total flow of the system.
In this sense, analytical results of two segments with a common 
non-acceleration rate, (4)-(6), showing the existence of
the ``quantized'' flux is the basis of all subsequent results showing the
metastable jamming blocks.

% % % % % % % % % %
% % % % % % % % % %
\section{Summary and Prospects}

In real-life traffic, we often experience different mode of congestion on different days
on same roads with similar number of cars.
In this article, we have shown that this puzzling feature is captured with
a very simple cellular automaton model.
Our model has a property of self-aggregating
block of jam.  This jam functions as a dynamical bottleneck, and leads to the existence 
of the third phase at intermediate density where visually striking non-unique flux 
is observed for a given density.  
It is our view, that among the multiple possible variants of original NS models,
ones with such property are definitely worth further attention.
On the other hand, we want to make clear that the introduction of
the alternative new rule (\ref{rule}) is not the prerequisite
for the existence of the third phase.
In fact, it can be shown through the numerical simulations 
that the three phase structure is already present 
in {\it multisegment model with standard NS rules}.  
With our modified rule that allows the formation of metastable blocks of jam, 
however, the new phase obtains distinct characteristics of discretized flux. 

The existence of metastable states has been already observed in several 
traffic flow models \cite{BS98, AS01, NI04} that effectively amount to introduce 
some extra rules to the standard Nagel-Schreckenberg model.  
In particular, early indication of discretized flux can be observed already in \cite{NI04}.
Our model is arguably the simplest of such extensions since it actually
{\it reduces} the number of rules.   
More important point than the smaller number of rules is that, with our model, 
we are able to show a clear connection between
the metastable states and the newly found intermediate phase
featuring the discretized flux.  In light of our results, it is suspected that the existence of
the metastable states observed in elaborate realistic traffic models do not
come from detailed model-specific assumptions but have a simple
dynamics behind it which is captured in our minimal model.

The formation of metastable clusters  is recently observed
in a cellular automaton model of ants' trail \cite{NC03}.  
It would be beneficial to compare our results with those of this work 
to gain a better insight into the dynamics of block of jam.

It is worth mentioning that the state of the system in the newly identified third regime
at intermediate density shares many common features with
so-called {\it self-organized criticality} \cite{BT87}.  It is as though the system is
at critical point  for extended range of parameter
values with various competing metastable states formed spontaneously without
fine-tuning of the system.
It should be interesting to check, for example,
the existence of 1/f type fluctuation on some measure, 
a hall mark of self-organized criticality,
to further clarify the connection.
\\

%We would like to thank Prof. K. Takayanagi and Prof. T. Kawai
%for enlightening discussions.
%
%\newpage
%--------------------------%

\end{document}